\documentclass[twocolumn,showpacs,preprintnumbers,amsmath,amssymb,footinbib,longbibliography,superscriptaddress]{revtex4-1}
\usepackage[utf8]{inputenc}
\usepackage[english]{babel}
\usepackage{lmodern}
\usepackage{amsmath}
\usepackage{graphicx}
\usepackage{physics}
\usepackage{amssymb}
\usepackage{epstopdf}
\usepackage{ upgreek }
\usepackage{color}
\usepackage{setspace}
\usepackage{bbm}
\usepackage{ mathrsfs }
\usepackage[bottom]{footmisc}

\graphicspath{{Figures/}}

\usepackage[toc,page]{appendix}
\usepackage{tocloft}

\definecolor{azul_ri}{RGB}{0,102,179}

\usepackage{hyperref}
\definecolor{darkblue}{rgb}{0.031,0.282, 0.49} 
\hypersetup{
    colorlinks      = true,
    linkcolor       = darkblue,
    menucolor       = black,
    citecolor       = darkblue,
    urlcolor        = blue,
}

\newcommand{\figref}[1]{\mbox{Fig.~\ref{#1}}}

\newcommand{\secref}[1]{\mbox{Section~\ref{#1}}}

\renewcommand{\eqref}[1]{\mbox{Eq.~(\ref{#1})}}

\newcommand{\figpanel}[2]{Fig.~\hyperref[#1]{\ref*{#1}(#2)}}
\newcommand{\figpanels}[3]{Fig.~\hyperref[#1]{\ref*{#1}(#2)-(#3)}}
\newcommand{\figpanelNoPrefix}[2]{\hyperref[#1]{\ref*{#1}(#2)}}
\newcommand{\figlink}[2]{\hyperref[#1]{\color[rgb]{0.031,0.282, 0.49}#2}}

\begin{document}

\title{Successive quasienergy collapse and breakdown of photon blockade in the few-emitter limit}

\author{T.~Karmstrand}
\author{G.~Johansson}
\affiliation{Department of Microtechnology and Nanoscience (MC2), Chalmers University of Technology, 412 96 Gothenburg, Sweden}
\author{R.~Guti\'{e}rrez-J\'{a}uregui}
\email[Email:]{rg-jauregui@fisica.unam.mx}
\affiliation{Departamento de F\'{i}sica Cu\'{a}ntica y Fot\'{o}nica, Instituto de F\'{i}sica, Universidad Nacional Aut\'{o}noma de M\'{e}xico, Ciudad de M\'{e}xico, 04510, M\'{e}xico}

\begin{abstract}
The emergent behavior that arises in many-body systems of increasing size follows universal laws that become apparent in order-to-disorder transitions. While this behavior has been traditionally studied for large numbers of emitters, recent progress allows for the exploration of the few-emitter limit, where correlations can be measured and connected to microscopic models to gain further insight into order-to-disorder transitions. We explore this few-body limit in the driven and damped Tavis--Cummings model, which describes a collection of atoms interacting with a driven and damped cavity mode. Our exploration revolves around the dressed states of the atomic ensemble and field, whose energies are shown to collapse as the driving field is increased to mark the onset of a dissipative quantum phase transition. The collapse occurs in stages and is an effect of light-matter correlations that are overlooked for single atoms and neglected in mean-field models. The implications of these correlations over the macroscopic observables of the system are presented. We encounter a shift in the expected transition point and an increased number of parity-broken states to choose from once the ordered phase is reached.  

\end{abstract}
\maketitle

\section{Introduction}


A collection of atoms coupled to a single radiation mode becomes correlated as field and matter exchange excitations. The exchange rate is determined from the interference of individual transition paths and, as such, depends on both the number of photons and atoms, and how they are arranged in space. In practical situations where this composite light-matter system couples to an environment, its coherent coupling is rivaled by dissipation. Dissipation causes a profound change in the behavior of the system that goes beyond losing excitations. For example, atoms that radiate collectively---as found in dense arrays coupled to free-space or a bad cavity---become locked in phase as they fall down the excitation ladder~\cite{Gross_1982,Carmichael_1999,Masson_2022}. This locking leads to a short, superradiant burst that is in contrast to the exponential decay found when atoms radiate individually or the oscillations they experience when there are no decay paths~\cite{Brune_1996,Meekhof_1996,Wallraff_2004,Fink_2009}. These different responses can be traced back to three different ways the environment divides this quantum system into separate components.

Separability is at the center of quantum optics studies. It allows for individual properties to be assigned, and measured, to different parts of a composite quantum system~\cite{Noh_2010}. It also plays a role in defining two of its regimes of research: a strong coupling regime where coherence dominates over dissipation; and a weak coupling one where dissipation overwhelms coherence. The former favors experiments with low levels of excitation, while the latter favors high levels. This practical gap separating both regimes translates to theoretical descriptions and technological applications. For example, the correlations that arise between different parts of a composite system in strong coupling can be tracked for low levels of excitation and then leveraged to test fundamental quantum behavior as formation and loss of entanglement~\cite{Haroche_2013,Wineland_2013,Haroche_2020} or perform computational tasks~\cite{Monroe_1995,Schuster_2007,Devoret_2013,Daiss_2021}. Weak coupling conditions, in turn, justify the approximations that make analytical calculations tractable in large quantum systems. The response of a laser turning into action~\cite{Haken_1970,Rice_1994} or microscopic models of optical devices~\cite{Bonifacio_1978,Drummond_1981,Reid_1985} are usually described using a system-size expansion where correlations are neglected up to a given order.


In this paper, we explore strong-coupling effects over few emitters and high levels of excitation. Our work is motivated by recent experiments where increasingly large quantum systems are built from individual components~\cite{Browaeys_2020,Kaufman_2021}. The experiments rely on a control over individual emitter positions that allows to construct arbitrary patterns. For dense and ordered patterns, the emitters display a collective response already made available at small numbers, as has been shown for neutral atoms~\cite{Leseleuc_2019,Rui_2020}, molecules~\cite{Anderegg_2019,Holland_2023}, ions~\cite{Katz_2023}, and artificial atoms~\cite{Kannan_2020,Wang_2020,Zanner_2022}. We are particularly interested in experiments where these arrays are placed inside a high-quality cavity~\cite{Fink_2009,Reimann_2015,Casabone_2015,Begley_2016,Neuzner_2016,Liu_2023,Masson_2023}. The versatility brought by emitter arrays brings attention to a {\textit{few-body}} limit. A limit where cooperative behavior arises while correlations are still tractable. 


To explore strong-coupling effects in this few-body limit, we consider the driven and damped Tavis--Cummings (TC) model, which describes the ideal coupling between a collection of two-level emitters and a single coherently driven mode of a lossy cavity. The driving field allows us to move between different levels of excitation and shift our focus from the transient behavior of superradiance to the steady-state behavior characteristic of systems driven out of equilibrium. We connect with previous results by showing that, for weak driving fields, the cavity remains near its vacuum state as the atoms block the absorption of photons. And then, by ramping up the field, show how this blockade is broken for few atoms. We show that---unlike the breakdown found for single atoms~\cite{Alsing_1990,Alsing_1992,Carmichael_2015,Fink_2017,GJ_2018,Lang_2011,Vukics_2019,Chen_2019,Mavrogordatos_2022} or the extension supported for many atoms in mean-field~\cite{GJ_2018a}---the breakdown occurs in stages, with different excitation paths being opened at discrete intervals. Each path follows a partial collapse of the quasienergies of the driven TC model, where the discrete spectrum found in the absence of a driving field gives way to a continuous one. The fact that this organization occurs successively offers a striking illustration of light-matter correlations over macroscopic observables and their effect over cooperative phenomena as parity-breaking. It is a strong-coupling effect in the few-body limit, which helps to bridge the gap between the single and the many.

The paper is organized as follows. We begin in Sec.~\ref{Sec:collapse} by introducing the driven Tavis--Cummings model. We find the quasienergy spectrum of this model numerically for small atom numbers to motivate the study of cooperative effects in many atoms. Then, in Sec.~\ref{Sec:cavity_loss}, we consider the effect of dissipation via cavity loss. Signatures of cooperativity are shown in the steady-state photon number and Wigner distribution of the cavity field. The latter displays a multistable response, which we relate to excitation channels that are being opened after each quasienergy collapse. The effect of quantum fluctuations is further studied in Sec.~\ref{Sec:stochastic}. Fluctuations appear to mold the behavior of the system around the closed-system dressed states, while trying to separate from the environment. Section~\ref{Sec:Liouville} is left to describe changes in the system via the Liouvillian spectrum. We conclude in Sec.~\ref{Sec:Summary} with a summary and future perspectives.

\section{Collapse of the driven Tavis--Cummings model}\label{Sec:collapse}

The ability to couple collections of artificial~\cite{Fink_2009} or neutral atoms to a single cavity mode~\cite{Liu_2023,Masson_2023} allowed to measure strong-coupling effects using an increasing atomic number. The experiments showcased an incredible control over atomic positions joined by an ability to probe their collective state via coherent driving fields. While the effect of the drive is negligible for these experiments, it has a profound effect when increased~\cite{Carmichael_2015,Fink_2017}. We begin by exploring these effects in the absence of dissipation. 

\subsection{Driven Tavis--Cummings model}

We consider a collection of $n_{at}$ atoms placed inside a driven cavity. Each atom is modeled as a two-state system of resonance frequency $\omega_{0}$ that is coupled to a dominant cavity mode of the same frequency. The mode is driven in resonance by a coherent source of amplitude $\varepsilon$. Atoms and cavity interact via a coupling strength $\Omega(\mathbf{r}_{m})$ that depends on the local profile of the mode evaluated at the atomic position $r_{m}$, which we assume can be controlled deterministically. When all atoms share the same coupling strength, the system is accurately described by the driven Tavis--Cummings model
\begin{align}\label{eq:TC-Hamiltonian_drive}
\mathcal{H}_{\varepsilon} = &\hbar \omega_{0} \left( \tfrac12 \hat{S}_{z} + \hat{a}^{\dagger}\hat{a} \right) + \hbar \Omega \left(\hat{S}_{+}\hat{a} + \hat{S}_{-} \hat{a}^{\dagger} \right) \nonumber \\
+& \hbar \varepsilon \left( \hat{a}e^{i\omega_{0}t} +  \hat{a}^{\dagger}e^{-i\omega_{0}t} \right) \, .
\end{align}
Here $\hat{a}$ and $\hat{a}^{\dagger}$ are the creation and annihilation operators for the cavity mode while $\hat{S}_{z}$ and $\hat{S}_{\pm}$ are the collective operators for the atoms. The collective operators can be written in terms of individual operators $\hat{\sigma}_{\pm}^{(m)}$ as 
\begin{equation}\label{eq:collective_operators}
\hat{S}_{\pm} = \sum_{m=1}^{n_{at}} \hat{\sigma}_{\pm}^{(m)} \, ,
\end{equation}
and obey the angular momentum commutation relations $\left[ \hat{S}_{+}, \hat{S}_{-} \right] = 2\hat{S}_{z}$, $\left[  \hat{S}_{z} , \hat{S}_{\pm},\right] = \pm \hat{S}_{\pm}$

The eigenstates and energies of $\mathcal{H}_{\varepsilon=0}$ have been computed in the absence of a driving field $(\varepsilon=0)$~\cite{Tavis_1968,Scharf_1970,Narducci_1973}. They are obtained by describing the collective atomic state as a spin of total angular momentum $s$ and projection $m_{s}$ 
\begin{equation}
\hat{S}_z\ket{s,m_s}=m_s\ket{s,m_s}
\end{equation}
that exchanges excitations with a cavity mode of photon number $n_{ph}$ 
\begin{equation}
\hat{a}^{\dagger}\hat{a}\vert n_{ph}\rangle=n_{ph} \vert n_{ph}\rangle\, .
\end{equation}
Since the total number of excitations $n$ is conserved, the eigenstates can be written within a bare basis $\vert m_{s},n_{ph}\rangle$
\begin{equation}\label{eq:dressed_states}
\vert \Psi_{n,\ell} \rangle = \sum_{p} c_{\ell;p} \vert p,n-p \rangle \, ,
\end{equation}
where $\ell$ determines a particular eigenstate inside a space of $n= n_{ph} + p$ excitations with $p=\tfrac12 n_{at}+m_{s}$ excitations stored in the atoms~\cite{us_2023}. The amplitudes $c_{l;p}$ are given by the roots of a polynomial~\cite{Tavis_1968,Scharf_1970,Narducci_1973}. Throughout, we consider situations where the the total spin $s$ is conserved and omit the $s$-index when possible.

The states $\vert \Psi_{n,\ell} \rangle$ define the dressed states of the TC model. They describe the collection of atoms and cavity mode as a coupled entity. This coupling causes their energies to split symmetrically from the bare energy as
\begin{equation}
E_{n,\pm \ell} = n \hbar \omega_{0} \pm \hbar {\Omega}_{n, \vert \ell \vert } \, ,
\end{equation}
where the splittings ${\Omega}_{n, \vert\ell\vert}$ are also obtained as roots of a different polynomial~\cite{Scharf_1970}. We use $ \Omega_{n,\vert \ell \vert}$ for the frequency splittings and keep $\Omega$ for the vacuum Rabi frequency.

\subsection{Quasienergies of the driven Tavis--Cummings model}

The driving field causes a periodic modulation of the Hamiltonian $\mathcal{H}_{\varepsilon}$ at a frequency $\omega_{0}$. Moving to an interaction picture with respect to the free terms $\hbar \omega_{0}(\hat{S}_{z}/2 + \hat{a}^{\dagger}\hat{a})$, this explicit time-dependence is removed to give
\begin{equation}\label{eq:TC-driven_int}
\tilde{\mathcal{H}}_{\varepsilon} = \hbar \Omega (\hat{S}_{+} \hat{a} +  \hat{S}_{-}\hat{a}^{\dagger}) + \hbar \varepsilon (\hat{a} + \hat{a}^{\dagger}) \, .
\end{equation}
In so doing, attention is shifted from energies and eigenstates into quasienergies and time-periodic steady-states that satisfy
\begin{equation}\label{eigenvalue_eq}
\tilde{\mathcal{H}}_{\varepsilon} \vert \psi_{\alpha} \rangle = \tilde{E}_{\alpha}  \vert \psi_{\alpha} \rangle\, .
\end{equation}

The quasienergies $\tilde{E}_{\alpha}$ show a simplified spectrum in which the bare frequency $n \hbar \omega_{0}$ is removed. This allows us to unveil a structural change in the system as the driving field amplitude is increased, where parts of the discrete spectrum of the TC model collapse at critical drive amplitudes to become continuous above them. The transition between a discrete and a continuous spectrum reflects a fundamental change in the behavior of the system that is caused by the competition between a nonlinear coupling and a linear drive.

We examine this change by taking as a starting point the single atom results~\cite{Alsing_1990,Alsing_1992,Carmichael_2015,Fink_2017,Vukics_2019,GJ_2018,Lang_2011}, that, in our notation, are represented by $s=1/2$ and $m_{s}=\pm 1/2$. In the absence of a driving field, the response of a single atom coupled to a cavity is described by the Jaynes-Cummings dressed-states~\cite{Themis_2021}. These states are solutions to Eq.~(\ref{eigenvalue_eq}) and display a characteristic $\sqrt{n}$-spectrum. As the driving field is turned on, these states begin to spread along different excitation manifolds, transforming into
\begin{align}\label{eq:TC-dressed-states}
\vert \psi_{n,\pm 1/2} \rangle  = &\tfrac12 \hat{\mathcal{S}}(z) \hat{\mathcal{D}}(r)  \sqrt{1-\Xi}  (\vert g, n-1 \rangle \pm \vert e,n \rangle) \nonumber \\
\mp&\tfrac12  \hat{\mathcal{S}}(z) \hat{\mathcal{D}}(r) \sqrt{1+\Xi} (\vert g,n \rangle \pm \vert e,n-1 \rangle)
\end{align}
 where atomic ground $\vert g \rangle$ and excited states $\vert e \rangle$ carry zero and one excitation, respectively,  while
 \begin{equation}
\Xi = \left[1-(2\varepsilon/\Omega)^{2} \right]^{1/2} \, .
\end{equation}
$\hat{\mathcal{D}}(r)$ and $\hat{\mathcal{S}}(z)$ are displacement and squeezing operators that depend on the driving field through the parameters $r$ and $z$~\cite{Alsing_1990,Chen_2019}.
The quasienergies of $\vert \psi_{n,\pm 1/2} \rangle$, 
\begin{equation}\label{eq:TC-driven_int-energies}
\tilde{E}_{n,\pm 1/2} = \pm \sqrt{n} \hbar \Omega \left[1-(2\varepsilon/\Omega)^{2} \right]^{3/4} \, ,
\end{equation}
maintain the discrete $\sqrt{n}$-spectrum due to the destructive interference between two paths out of a each Fock state $\vert n_{ph} \rangle$. The paths are induced by the driving field or the atom, as revealed by inserting Eq.~(\ref{eq:TC-dressed-states}) into~(\ref{eigenvalue_eq}). The destructive interference of these paths, however, can not occur indefinitely. While the driving field can be manipulated externally, the atomic coupling is bounded by $\Omega$. Such that the atoms need to radiate more strongly as the field is increased to cancel its effect. This boundary is reflected in the separation between the  quasienergies, which begins to narrow. The narrowing continues up to a critical drive amplitude
\begin{equation}
\varepsilon_{\text{col}}(\tfrac12) = \tfrac12 \Omega  \, ,
\end{equation}
%
when the atom can no longer cancel the driving field and the separation closes~\cite{Alsing_1992,GJ_2018}. Equation~(\ref{eq:TC-dressed-states}) represents a separable product of field and atomic states at this point.

\begin{figure}
\begin{center}
\includegraphics[width=.95\linewidth]{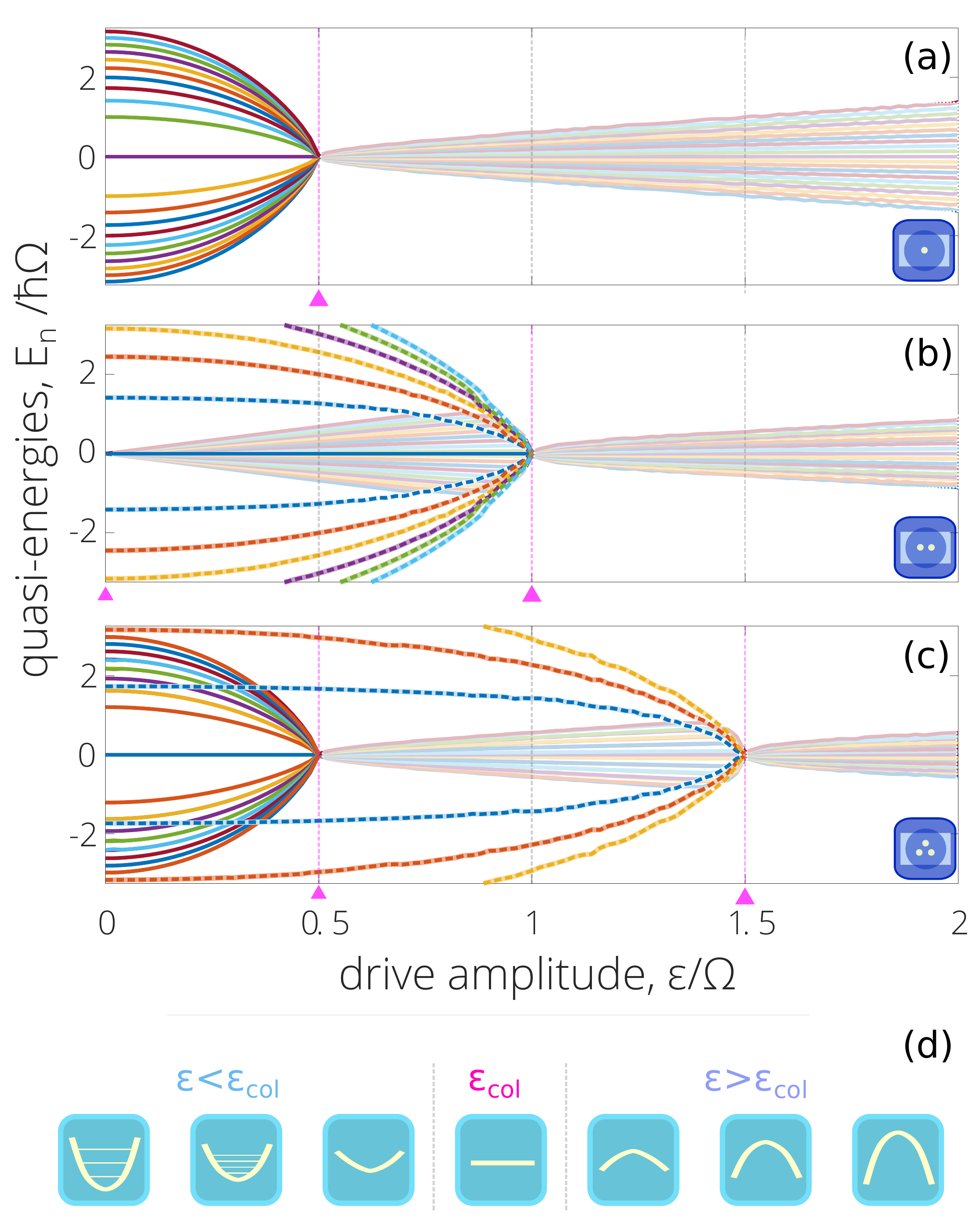}
\caption{(a-c) Quasienergies of the driven TC-Hamiltonian for $n_{at}= \lbrace 1,2,3 \rbrace$  obtained via numerical diagonalization of Eq.~(\ref{eq:TC-driven_int}). A discrete spectrum found at small drive amplitudes collapses at the critical drives $\varepsilon_{\text{col}}(m_{s}) = \Omega m_{s}$ with $m_{s}=0, (\tfrac12), 1, (\tfrac32), \ldots,\tfrac12 n_{at}$ for even (odd) number of atoms. After each collapse (pink triangles) a continuous spectrum rises (shaded lines). These partial collapses allow for states to be organized into separate groups denoted by the cooperative number $m_{s}$, drawn here as solid and dashed lines.  (d) This change from discrete to continuous spectrum is understood in analogy to a harmonic potential that is inverted by the driving field, as shown for one atom in Refs.~\cite{GJ_2018,Blas_2021}. } \label{Fig:collapses}
\end{center}
\end{figure}

The discrete quasienergies of the driven Jaynes-Cummings model collapse at this critical drive. The collapse then marks a point where discrete states with a well-defined quantum number $n$  give way to continuous ones spanning infinite $n$ above it. To explore the effect above the collapse, Guti\'{e}rrez-J\'{a}uregui and Carmichael build upon a formal analogy between the model and a relativistic particle trapped in a harmonic potential, already suggested by the $\sqrt{n}$-spectrum~\cite{GJ_2018}. The authors show how the harmonic potential is smoothened for weak drives, flattened at the critical point, and, ultimately, inverted afterwards. Thus explaining the narrowing separation between the quasienergies and their behavior after the collapse, as sketched in \figpanel{Fig:collapses}{d}. 

The destructive interference between the field radiated by the atoms and the one introduced by the drive is at the center of the quasienergy collapse for a single atom. A similar mechanism should follow for many atoms, where cavity and atoms exchange excitations at an accelerated rate that not only depends on the total excitation number $n$, but on how the excitations are distributed among them [see, e.g., the dressed-states of Eq.~(\ref{eq:dressed_states})]. With atoms and cavity exchanging excitations at different rates for each dressed-state, the collective system is expected to organize into groups that collapse at different drive amplitudes.

We obtain the amplitudes at which these collapses occur from the state ansatz
\begin{equation}
\vert \psi_{\nu} \rangle =  \hat{\mathcal{S}}(z_{\nu}) \hat{\mathcal{D}}(r_{\nu}) \vert \phi_{\nu} \rangle \, ,
\end{equation}
where the index $\nu$ encompasses all quantum numbers. The ansatz is obtained in analogy to the single atom case, where displacement and squeezing operators allow for the eigenstates to spread along different excitation numbers as the driving field is increased. But, in order to capture the different rates at which excitations are  collectively exchanged between field and atoms, we allow for the parameters $z_{\nu}$ and $r_{\nu}$ to depend on $\nu$. Under this ansatz the eigenvalue equation~(\ref{eigenvalue_eq}) is taken into
%
\begin{equation}\label{eq:collapse_points-1}
\hbar\left( e^{-z_{\nu}}(\hat{a} + \hat{a}^{\dagger}) - 2r_{\nu} \right) \left(\varepsilon + \Omega \hat{S}_{x} \right) \vert \phi_{\nu} \rangle = \tilde{E}_{\nu} \vert \phi_{\nu} \rangle \, .
\end{equation}
Equation~(\ref{eq:collapse_points-1}) can be expanded within the bare basis to construct a system of equations that connects atomic states carrying different numbers of excitations to identify the interfering paths. But, as the atoms can hold more than one excitation collectively, these equations can not be reduced to those of a harmonic oscillator following the methods of Ref.~\cite{Alsing_1992}. Equation~(\ref{eq:collapse_points-1}), however, can be solved at the drive amplitudes 
\begin{equation}\label{eq:collapse_points-2}
\varepsilon_{\text{col}}(m_{s}) = \Omega \vert m_{s} \vert \, ,
\end{equation}
where the separable states
\begin{equation}\label{eq:collapse_state}
\vert \phi_{\nu} \rangle = e^{-i \pi \hat{S}_{y}/2} \vert s, -\vert m_{s}\vert  \rangle \otimes \sum_{n_{ph}} c_{nph}^{(\nu)} \vert n_{ph} \rangle
\end{equation}
all have zero quasienergy. Atoms inside these states are described collectively by a cooperative number $m_{s}$, which takes the values $m_{s}=s,s-1,\dots,-s$ with $s = n_{at}/2$. Their collective state is rotated by $\pi/2$ along the $S_{y}$-axis, causing the atoms  to radiate at their maximum possible strength for a given $m_{s}$ to cancel the incoming drive~\cite{Haroche_Raimond_2006}. At this point the state of the field is independent of the atomic state, giving way to the large degeneracy that characterizes the collapse. Past the critical drive $\varepsilon_{\text{col}}(m_{s})$ the atoms can no longer cancel the incoming field, which causes a continuous quasienergy spectrum.

Figure~\ref{Fig:collapses} shows the quasienergies that are closer to zero in the driven TC model for $n_{at}=1,2,3$. The results exemplify the transition from a discrete to a continuous spectrum as a function of drive strength, with partial collapses occurring at the points predicted by Eq.~(\ref{eq:collapse_points-2}) and marked by pink triangles in the Figure. We plot the twenty quasi-energies that are closer to zero and, according to their collapse point, identify their cooperativity $m_{s}$. For clarity, we extend the lines corresponding to states of large cooperativity to low drives in the case of two and three atoms ($m_{s}=1 $ and $3/2$ respectively) and draw them as dashed lines. In both cases the spectrum shows crossings between states of different cooperativity, see \textit{e.g.}, $\varepsilon \simeq 0.3-0.4 \Omega$ for three atoms, suggesting further that states of different cooperativity remain decoupled as the driving field is increased. All lines are obtained by numerical diagonalization using a truncated Fock-state basis of $n_{\text{ph}}=500(n_{\text{at}}+1)$. The small oscillations seen for two and three atoms are a numerical artifact caused by imperfect sorting of these solutions.

\section{Cavity loss and coexisting states}\label{Sec:cavity_loss}

The successive collapses illustrate how atoms and cavity organize to minimize energy-costly fluctuations. They settle into dressed states that reduce fluctuations in the excitation number (characterized by $\Omega$) before spreading along an infinite number of excitations to minimize fluctuations in the $(\hat{a}+\hat{a}^{\dagger})$-quadrature (characterized by $\varepsilon$) as the drive is increased. Dissipation changes this picture in a profound way. Understood as the effect of a surrounding environment monitoring the system, dissipation introduces decay channels that balance the driving field and nonlinear coupling to relax the system into a steady state. The particular form of the decay channels determines how the system is separated by the environment. The system behavior is no longer settled by minimizing energy-fluctuations only, but also the information exchange with the environment.

We consider dissipation in our model via cavity loss at a rate $\kappa$. With the inclusion of this decay channel, the master equation for the density matrix of the system $\hat{\rho}$ is 
\begin{align}\label{eq:master_2}
\dot{\hat{\rho}} = (i\hbar)^{-1}\left[ \tilde{\mathcal{H}}_{\varepsilon}, \hat{\rho} \right] + \kappa \mathcal{L}_{\hat{a}}\hat{\rho} \, ,
\end{align}
where loss is accounted by the Lindblad superoperator $$\mathcal{L}_{\hat{a}} \cdot \equiv 2 \hat{a}\cdot\hat{a}^{\dagger} - \cdot\hat{a}^{\dagger}\hat{a} - \hat{a}^{\dagger}\hat{a}\cdot.$$ 
Equation~(\ref{eq:master_2}) defines the driven Tavis--Cummings model with dissipation. In the following sections we present numerical solutions to this equation in the strong coupling regime. The solutions are obtained using a Runge-Kutta method until a steady-state is reached. They highlight the different roles dissipation can play over the state of the system and its macroscopic observables.

\subsection{Photon number expectation and atomic population}

\begin{figure}
\begin{center}
\includegraphics[width=\linewidth]{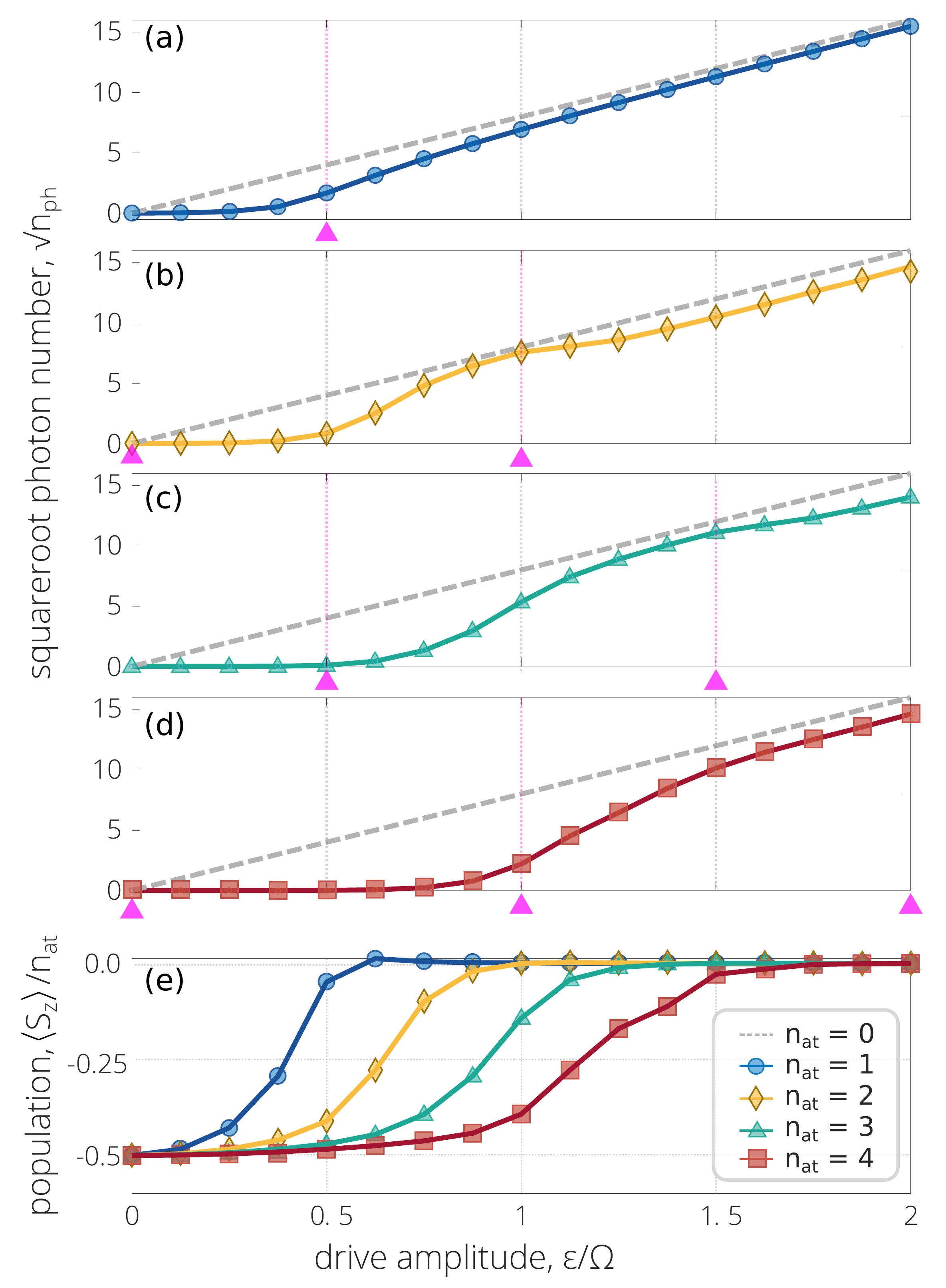}
\caption{Photon number expectation and atomic population in the steady state for $n_{at}= \lbrace 1,2,3,4 \rbrace$ atoms and $\Omega = 8 \kappa$. At weak drives, the nonlinear coupling between atoms and cavity leads to a photon blockade that is reflected in a vanishing cavity field. As the drive is increased and atoms begin to saturate, excitation paths that connect ground and excited states are opened to break this blockade. The paths open near each collapse point of Fig.~\ref{Fig:collapses}, marked by pink triangles. When a new path is opened there is a slowdown in the photon number caused by an average between low and high-excitation paths. The response of a Lorentzian cavity (dashed lines) is recovered when all paths are opened and atoms fully saturate.  } \label{Fig:photon_number}
\end{center}
\end{figure}

Figure~\ref{Fig:photon_number} shows the square root of the photon number expectation in the steady state,
\begin{equation}
\bar{n}_{ph}^{1/2} = \langle  \hat{a}^{\dagger} \hat{a} \rangle_{ss}^{1/2}      \, ,
\end{equation}
as a function of drive strength for different atom numbers with individual coupling strength $\Omega = 8 \kappa$. We plot the square root to highlight the regular intervals at which collapses occur while allowing for comparison with a Lorentzian cavity whose photon number expectation is $\bar{n}_{ph} = (\varepsilon/\kappa)^{2}$.  

The results are obtained for increasing atom numbers to  unveil trends in the steady-state behavior. Consider first the single atom case (blue circles). Here, the cavity displays a vanishing photon number for weak drives that begins to rise around the collapse point $\varepsilon_{col}(1/2)$. Attention should be given to the slope of the curve, which increases past this point before it saturates to $\varepsilon /\kappa$ at large drives (dashed line). This behavior is to be compared with the cases for two (yellow diamonds) and three atoms (green triangles). In both cases, the cavity displays a vanishing photon number for weak drives---a behavior that remains for stronger drives as the atom number increases---and saturates to the Lorentzian cavity past the last collapse [found at $\varepsilon_{col}(1)$ for two atoms and at $\varepsilon_{col}(3/2)$ for three]. In the region between the limits of vanishing photon number and fully saturated cavity, an increasing photon number is seen to {plateau}---or slow down---near the collapse points.

We explain these changes in photon number by introducing excitation ladders that are opened near each collapse point. The ladders are built from the TC dressed states of Eq.~(\ref{eq:dressed_states}). These states display a complicated anharmonic energy spectrum at low excitation $(n \lesssim n_{at})$ that begins to regularize as more excitations are added. At high levels of excitation $(n \gg n_{at})$, the states  without drive are approximated by~\cite{Agarwal_1977}
\begin{align}\label{eq:dressed_states_broken}
\vert \Psi_{n,\ell=\pm m_{s}} \rangle &\simeq e^{-i \pi \hat{S}_{y}/2} \vert m_{s},  n-m_{s}- \tfrac12 n_{at} \rangle \, ,
\end{align}
with quasienergies
\begin{equation}\label{eq:quasi_energies_broken}
\bar{E}_{n,\ell=\pm m_{s}} \simeq  \pm 2 \hbar \Omega m_{s} \sqrt{n} \, .
\end{equation}
The driving field connects these excited states to the ground state via multi-photon resonances. Each step up the transition ladders is detuned by $\tilde{E}_{n+1,\ell}-\tilde{E}_{n,\ell}$, an intensity-dependent detuning that, at the high excitations implied by~(\ref{eq:quasi_energies_broken}), is brought close to resonance as
\begin{equation}\label{eq:det}
\tilde{E}_{n+1,\ell=\pm m_{s}}-\tilde{E}_{n,\ell=\pm m_{s}} \simeq  \pm \hbar \Omega m_{s} /  \sqrt{n} \, .
\end{equation}
This dependence on intensity marks a significant difference between low and high levels of excitation. In the former, the absorption of one photon brings a dressed state out of resonance with its neighbouring state and thus block the absorption of a second photon. In the latter,  multiple absorptions are likely.

Highly-excited dressed-states then present the system with harmonic ladders for the photon number to climb up. The ladders are distinguished by their cooperativity number $m_{s}$ and each one of them can be ascended via its upper $+m_{s}$ or lower $-m_{s}$ rungs. Strong drives are needed to move past the uneven steps found at low excitations---with increasing strength for larger $m_{s}$ and $n_{at}$---but, once the excited states are reached, transitions along each rung are tuned near resonance. This tuning is accompanied by a sharp change in photon number expectation. The intervals where the photon number increases sharply in Fig.~\ref{Fig:photon_number} correspond to regions where ladders connect ground and excited states. The slowdown experienced near subsequent collapse points shows the emergence of additional coexisting ladders. Since newly emergent ladders are slightly out of resonance at first, they reach lower excitation levels. This is reflected as a slowdown in photon number when averaged over the contributions of all possible ladders. 

The sharp change in photon number is a strong-coupling effect. The decay rate $\kappa$ gives the excited states a linewidth that allows for higher levels to be excited even when slightly out of resonance. The sharp changes are thus smoothened over a region. As we move deeper into the strong-coupling regime $(\Omega \gg \kappa)$, this sharpness is recovered and eventually leads to a dissipative phase transition. Single-atom results~\cite{Carmichael_2015,Fink_2017} and mean-field models for many atoms~\cite{GJ_2018a} predict such a transition to occur at the last collapse $\varepsilon_{\rm col}(n_{at}/2)$. The mean-field models, in particular, neglect the correlations that form between cavity and atom. They lead to a collective atomic state that radiates as a single, large dipole of coupling strength $n_{at} \Omega$. The dipole cancels the incoming drive up to a critical drive strength after which the photon number begins to rise. The sharp change from empty to filled cavity is thus accompanied  by a change in the atomic population. 

We are interested in reconciling the picture of a sharp transition with that found above, where collapses occur gradually. For this reason, we plot the atomic population in Fig.~\ref{Fig:photon_number}. The population does not show the slowdown near the collapse points and its overall behavior resembles that found in Refs.~\cite{Carmichael_2015,Fink_2017,GJ_2018a}. The transition to a vanishing population becomes sharper as we delve deeper into the strong-coupling regime. The transition point, however, is displaced from the mean-field predictions. 

 In the next section, we show that each dressed-state ladder becomes more stable, as we move deeper into the strong-coupling regime. This stability is accompanied by sharp changes in the system as system dynamics in this regime are molded by fluctuations around the dressed states. The underlying structure that is dismissed in the mean-field approximation plays a significant role on the long-time behavior of the system. This suggests that if a sharp change in photon number is to be found for several atoms, it should arrive from an average over independent excitation ladders rather than a single superradiant one.

\subsection{Coexisting states and the multi-peaked Wigner distribution}

To further explore the excitation ladders, we obtain the steady-state Wigner distribution of the cavity field,
\begin{equation}
W_{ss}(\alpha, \alpha^{*}) = \frac{1}{\pi^{2}} \int  d^{2}x e^{\alpha x^{*} - \alpha^{*}x} \text{Tr}[\hat{\rho}_{ss} e^{x \hat{a}^{\dagger} - x^{*} \hat{a}}] \, .
\end{equation}
with $\hat{\rho}_{ss}$ the steady-state density matrix of the system.

\begin{figure}
\begin{center}
\includegraphics[width=\linewidth]{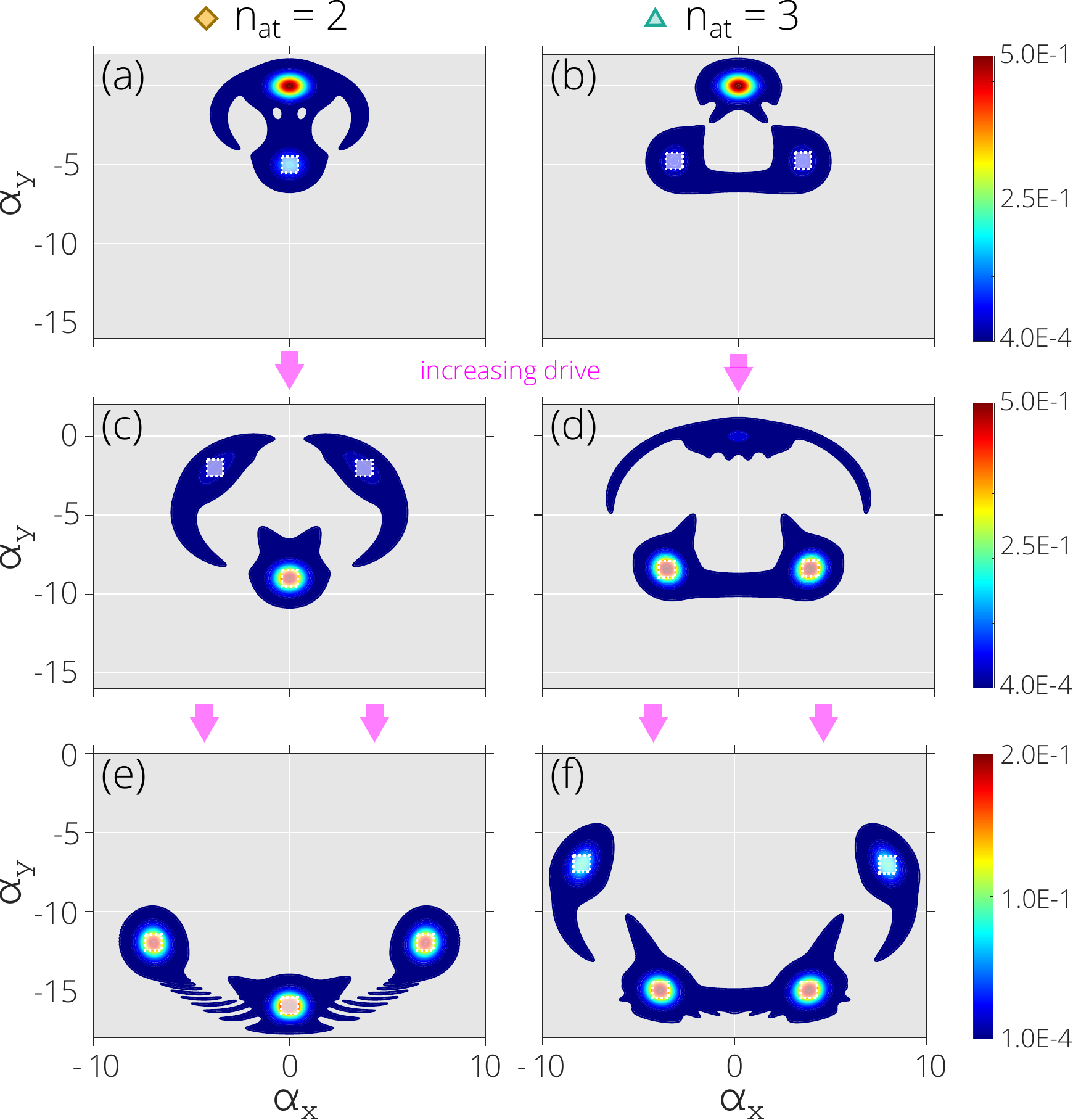}
\caption{Steady-state Wigner distributions for two and three atoms with increasing drive. As the drive is ramped up past each collapse point the distributions increase their modality, going from single to multi-modal. The peaks are centered around the mean-field prediction of independent excitation paths of Eq.~(\ref{eq:semiclassical-2}), marked by white squares. (a)-(f) Distributions $W_{ss}$ are plotted, respectively, for $\varepsilon/\kappa = \lbrace 5,7,9,10,16,16 \rbrace $ with $\Omega=8\kappa$ and $\alpha = \alpha_{x} + i \alpha_{y}$ .} \label{Fig:multi-peaked}
\end{center}
\end{figure}

Figure~\ref{Fig:multi-peaked} shows the steady-state Wigner distribution for two and three atoms subject to increasing drive strengths. The distributions go from single- to multi-peaked as the drive strength is increased past each collapse point. 
The peaks appear as doublets, displaced symmetrically from the $\text{Re}[\alpha]$-axis (plus an individual path for even atom numbers). Each pair signals two metastable solutions of equal phase that appear after a collapse. Their symmetry axis is imposed by the relative phase between the dipolar coupling and driving field~\cite{Curtis_2022}.

We associate the peaks to coexisting paths up the dressed-state ladders $m_{s}$. This picture is supported by a semiclassical model based on the mean-field equations for a single atom~\cite{Alsing_1990}. The equations treat the atom as a radiating dipole that spontaneously polarizes after a critical drive strength. For many atoms, we consider each cooperative response $m_{s}$ satisfies
\begin{subequations}\label{eq:semiclassical}
\begin{align}
&\dot{\alpha}_{ms} = -\kappa \alpha_{ms} - i \Omega \beta_{\pm ms} -i \varepsilon \, , \label{eq:field} \\
&\dot{\beta}_{ms} = i2\Omega \alpha_{ms} \xi_{\pm ms} \, , \\
&\dot{\xi}_{ms} = i\Omega (\alpha^{*}_{ms} \beta_{\pm ms} - \beta^{*}_{\pm ms} \alpha_{\pm ms}) \, ,
\end{align}
\end{subequations}
with $\alpha_{ms} = \langle \hat{a} \rangle_{ms}$, $\beta_{ms} = \langle S_{-} \rangle_{ms}$, and $\xi_{ms} = \langle S_{z} \rangle_{ms}$. The atomic population $\xi_{ms}$ and dipole moment $\beta_{ms}$ are connected by the conservation of total spin. We select $\vert \beta_{\pm ms} \vert^{2} +\xi_{\pm ms}^{2} = m_{s}^{2}$ to account for the cooperative response as that of a large dipole radiating at a  rate $m_{s} \Omega$. The steady-state solutions to these equations then undergo a bifurcation by balancing drive and dissipation. Below $\varepsilon_{col}(m_{s})$ an empty-cavity solution
\begin{equation}
\vert \alpha_{ms} \vert =0 \, ,
\end{equation}
is found. It gives way to two filled solutions
\begin{equation}\label{eq:semiclassical-2}
\vert \alpha_{\pm ms} \vert^{2} = (\varepsilon / \kappa)^{2} \left[1 - (m_{s}\Omega / \varepsilon)^{2} \right]\, ,
\end{equation} 
ith $\beta_{\pm ms} = \pm m_{s} \alpha_{\pm ms} / \vert \alpha_{\pm ms} \vert$, above this point. When substituted back into Eq.~(\ref{eq:semiclassical}), the filled-cavity solutions display two opposite phases that correspond to the plus and minus rungs of the $m_{s}$-ladder. The solutions are plotted as white squares in Fig.~\ref{Fig:multi-peaked}. They coincide with the peaks of the Wigner distribution that were obtained from a fully quantum approach. 

\section{Light-matter Correlations and parity-breaking states}\label{Sec:stochastic}
The picture of coexisting paths is useful to explain the photon number expectation and multi-peaked Wigner distributions shown above. This picture is built around the dressed states of Sec.~\ref{Sec:collapse} and single-atom results of Refs.~\cite{Alsing_1990}. Yet, little has been said about the role of the environment in mixing these states. We now include this effect explicitly to explain how the environment molds the steady state around the driven dressed states.

\subsection{Excitation-ladders as approximate pointer states}
\label{sec:pointer}

The environment acts on the system via the decay operators $\hat{a}$ of the master equation~(\ref{eq:master_2}). When acting on the dressed states without drive, these operators connect different $m_{s}$-ladders via
\begin{equation}
\hat{a} \vert \Psi_{s,n,\ell=m_{s}} \rangle = \sum_{\ell'=-s}^{s} c_{\ell \rightarrow \ell'}  \vert \Psi_{s,n-1,\ell'} \rangle  \, .
\end{equation}
At high excitations, where the dressed states are accurately described by Eq.~(\ref{eq:dressed_states_broken}), the likelihood to switch ladders follows from~\cite{Wiseman_1996,Carmichael_2008}
\begin{equation}
c_{\ell \rightarrow \ell '} =  \sum_{m_{a}=-s}^{s} d^{(s)}_{\ell,m_{a}}\left(\frac{\pi}{2} \right) d^{(s)}_{\ell',m_{a}}\left(\frac{\pi}{2} \right) \sqrt{n-\tfrac12 n_{at}-m_{a}} \nonumber \,,
\end{equation}
with $d^{(s)}_{m_{1},m_{2}}(\beta) = \langle s,m_{1} \vert e^{-i \beta \hat{S}_{y}}\vert s,m_{2} \rangle$ the Wigner matrix relating two angular momentum components~\cite{Messiah}.  The amplitudes $c_{\ell \rightarrow \ell'}$  represent de-excitation pathways whose amplitudes interfere constructively when moving within the same ladder $(\ell = \ell')$ and destructively otherwise $(\ell \neq \ell')$. This interference makes it more likely for the system to remain inside a particular $m_{s}$-ladder after a photon is radiated as can be shown explicitly for small atom numbers ($n_{at}= \lbrace 1,2,3 \rbrace$). For the case of two atoms the likelihood to switch is explicitly given by
\begin{align}
\hat{a} \vert \psi_{n+1,0} \rangle &= \sqrt{\frac{n^{2}-1}{n^{2}-\tfrac14}} \left( \frac{\sqrt{n}-\sqrt{n-1}}{2} \right) \frac{\vert \psi_{n,+}\rangle + \vert \psi_{n,-}\rangle}{\sqrt{2}} \, \nonumber \\
&+ \sqrt{\frac{n(n+1)}{n^{2}-\tfrac14}} \left( \frac{\sqrt{n}+\sqrt{n-1}}{2} \right)\vert \psi_{n,0}\rangle \, , \\
\hat{a} \vert \psi_{n+1,\pm 1} \rangle &= \sqrt{\frac{n}{n^{2}-1/4}}  \sum_{\ell=\pm 1} \left( \frac{2n\pm \ell \sqrt{4n^{2}-1}}{4} \right)\vert \psi_{n,\ell}\rangle \, \nonumber \\
&+ \sqrt{\frac{n-1}{n^{2}-1/4}} \frac{1}{\sqrt{8}}\vert \psi_{n,0}\rangle \, ,
\end{align}
making it more likely to remain inside a single ladder as the excitation number $n$ is increased.

This decreasing likelihood makes it possible to create approximate eigenstates of the jump operator within a single ladder
\begin{equation}\label{eq:pointer-1}
\hat{a} \sum_{n} c_{n} \vert \psi_{s,n,\ell=m_{s}} \rangle \simeq  \lambda \sum_{n} c_{n} \vert \psi_{s,n,\ell=m_{s}} \rangle \, .
\end{equation}
These approximate pointer-states minimize the rate at which a system loses its purity to the environment~\cite{Carvalho_2001}. The states correspond to the peaks in the Wigner distributions plotted in Fig.~\ref{Fig:multi-peaked}. The exact solutions for one~\cite{Carmichael_2015} and two atoms show that the effect of the jump operator approximates to $ \hat{a} \vert \psi_{s,n, ms} \rangle = \sqrt{n} \vert \psi_{n-1, ms} \rangle$ at large photon numbers $(n_{ph} \gg 1)$. The pointer states of Eq.~(\ref{eq:pointer-1}) then resemble coherent states with field amplitude $\alpha_{m_{s}}$ defined in Eq.~(\ref{eq:semiclassical-2}), thus explaining the success of this approximation to the peaks of the Wigner distribution. 

Since approximate pointer states can be created within a given $m_{s}$-ladder, the state of the driven TC-model will tend to settle inside one of these solutions before quantum fluctuations expel it towards another one. Switching ladders become less and less likely as we move deeper into the strong coupling regime $(\Omega/ \kappa \gg 1)$ and increase the driving strength $(\varepsilon/\kappa \gg 1)$  while keeping the rate $\varepsilon / \Omega$ constant. These parameters define a thermodynamic limit, a limit of large excitation numbers where the effect of fluctuations can be neglected. 
\begin{figure}
\begin{center}
\includegraphics[width=\linewidth]{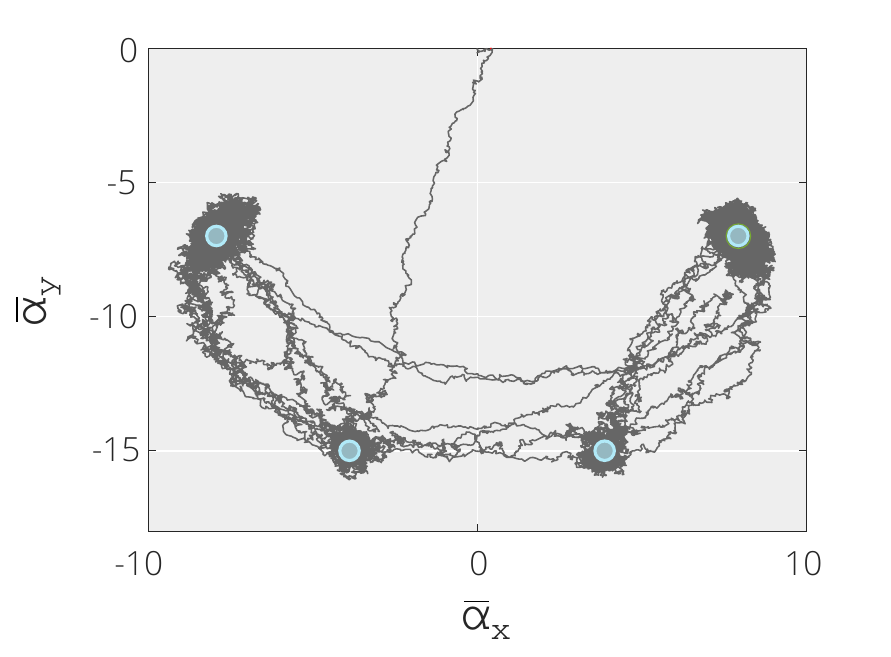}
\caption{Sample heterodyne current, $\bar{\alpha}_{t}= \bar{\alpha}_{x} + i \bar{\alpha}_{y}$, used to track the stochastic path followed by a pure state $\vert \psi_{t} \rangle$ via Eqs.~(\ref{eq:sto_sch})-(\ref{eq:sto_sch_3}). Fluctuations first drive the system from the ground state to a excited state via a particular $m_{s}$-ladder, a parity-breaking descision made over all possible states. As time advances, fluctuations cause the system to switch between parity-broken states [centered at blue dots obtained from the mean-field prediction of Eq.~(\ref{eq:semiclassical-2})]. The record is measured during a time interval of $\tau = 800 \kappa^{-1}$ with a filter bandwidth of $\kappa_{\tiny{\text{filt}}} = \kappa$ and the parameters of Fig.~\ref{Fig:multi-peaked}f.  } \label{Fig:heterodyne}
\end{center}
\end{figure}

\subsection{State switching in a stochastic evolution}

To present the path a pure state $\vert \psi_{t}\rangle$ would follow under the action of quantum fluctuations, we use a stochastic Schr\"{o}dinger equation corresponding to the master equation~(\ref{eq:master_2}):
\begin{equation}\label{eq:sto_sch}
d \vert \psi_{t} \rangle = \left[ (i \hbar)^{-1} \left( \mathcal{H}_{\varepsilon} -i\hbar \kappa \hat{a}^{\dagger} \hat{a} \right) dt  + \sqrt{2\kappa} \hat{a} dq \right] \vert \psi_{t} \rangle \, ,
\end{equation}
This stochastic evolution is understood as a quantum trajectory unravelling under heterodyne detection~\cite{Carmichael_2008}. The state of the field is readily measured via its output 
\begin{equation}\label{eq:sto_sch_2}
dq = \sqrt{\kappa} \frac{\langle \psi_{t}\vert \hat{a} \vert \psi_{t} \rangle}{\langle \psi_{t}\vert \psi_{t} \rangle}  dt + dZ
\end{equation}
with $dZ$ a complex Wiener increment with covariances $\overline{dZdZ} = \overline{dZ^{*}dZ^{*}} = 0$ and $\overline{dZdZ^{*}}=dt$~\cite{Carmichael_2008,Gardiner_1991}. The output is then filtered to simulate a heterodyne current record,
\begin{equation}\label{eq:sto_sch_3}
d\bar{\alpha}_{t} =  \tfrac12 \kappa_{\tiny{\text{filt}}} \left[ \bar{\alpha}_{t}dt - \sqrt{\kappa} dq \right],    
\end{equation}
commonly measured in experiments monitoring artificial atoms~\cite{Campagne-Ibarcq_2016}. 

We plot a sample record in Fig.~\ref{Fig:heterodyne} using the same parameters as Fig.~\ref{Fig:multi-peaked}f. Here, cavity and atoms were originally prepared in their ground state, from which they are quickly brought into one of the coexisting $m_{s}$-states. The particular state that is reached is arbitrary. It is a parity-broken state chosen by fluctuations over all possible states. The decision is revealed to the outside observer via the measurement record. Unlike the standard case of a single atom, atomic correlations create several subspaces for the system to reach [plotted as blue circles obtained from the mean-field prediction~(\ref{eq:semiclassical-2})]. Each pair corresponds to a parity-broken state, but the number of overall states depends on the atomic number. As we move closer to the thermodynamic limit, the ladder switching occurs less frequently.

\section{Liouvillian spectrum}\label{Sec:Liouville}
\begin{figure*}
\begin{center}
\includegraphics[width=1.\linewidth]{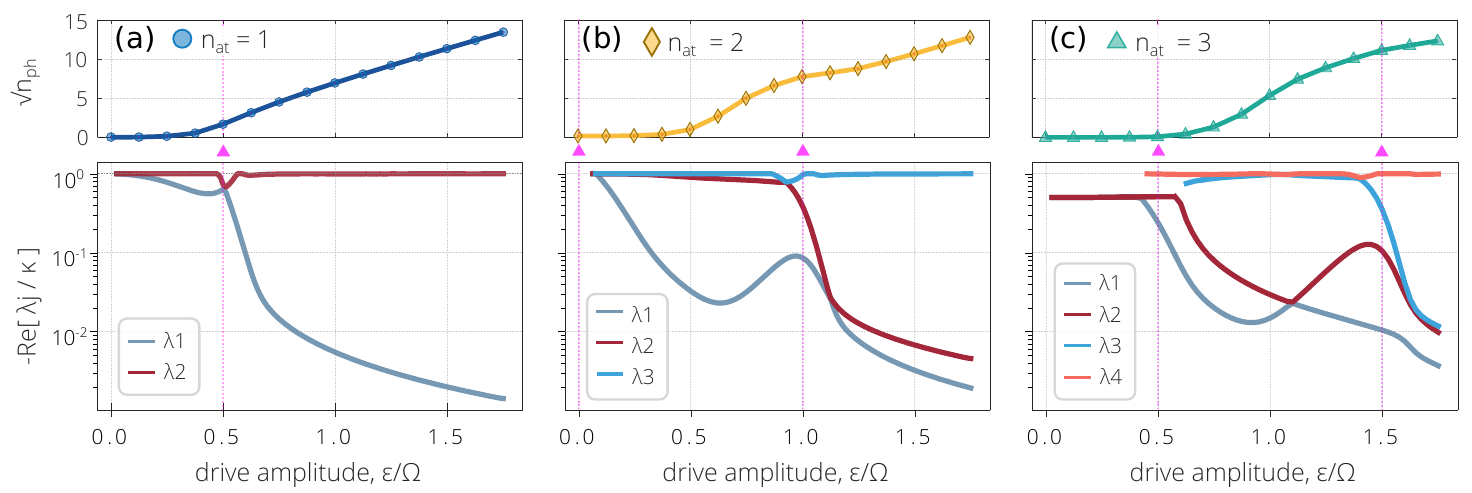}
\caption{Photon number and real part of the lowest eigenvalues $\lambda_{i}$ of the Liouvillian for $n_{at} = \lbrace 1,2,3 \rbrace$ atoms and $\Omega=8\kappa$. The distance between the smallest eigenvalues of the Liouvillian begins to close near the collapse points (pink triangles). This closing gap is accompanied by a slowdown in the photon number explained above as the emergence of new excitation ladders.} \label{Fig:liouvillian}
\end{center}
\end{figure*}

The correlations that emerge between atoms and cavity field lead to successive collapses in the absence of cavity decay and to the creation of multiple parity-broken states in its presence. These states are connected by fluctuations due to finite-system-size effects present in our numerical experiments. We have moved deeper into the strong-coupling regime to look for trends in the behavior of the system as we increase its size and approach the thermodynamic limit of Sec.~\ref{Sec:stochastic}. In this final section, we introduce the Liouvillian superoperator to characterize these effects. In close analogy to the Hamiltonian of Sec.~\ref{Sec:collapse}, the structure of the Liouvillian contains information about the system dynamics, such as relevant time-scales to reach its steady state, underlying symmetries, and the available space of steady states~\cite{Albert_2014}.  

The Liouvillian superoperator $\bar{\mathcal{L}}$ corresponding to the master equation~(\ref{eq:master_2}) is written explicitly as
\begin{equation}\label{eq:liouvillian}
	\bar{\mathcal{L}}\cdot= (i \hbar)^{-1}\comm{\tilde{\mathcal{H}}_{\varepsilon}}{\cdot} + \kappa(2\hat{a}\cdot\hat{a}^\dagger -\hat{a}^\dagger\hat{a}\cdot-\cdot\hat{a}^\dagger\hat{a}) \, .
\end{equation}
Its eigenvalues $\lambda_i$ and right eigenoperators $\rho_i$ are defined via the relation
\begin{equation}\label{ec:eigenvalue_Liouville}
	 \bar{\mathcal{L}}\rho_i, = \lambda_i \rho_i \, .
\end{equation}
The eigenvalues are complex. Their real components represent the rates of relaxation towards the steady state~\cite{Breuer_2010} and satisfy $\mathcal{R}e[\lambda_i]\leq 0$. Conservation of probability requires that at least one eigenvalue has a zero real part $\mathcal{R}e[\lambda_{0}]=0$. Its corresponding eigenoperator $\rho_0$ defines a steady state by $\hat{\rho}_{ss}\equiv \rho_0/\Tr[\rho_0]$. 
Being approximate eigenstates of jump operators and the Hamiltonian, we now explore whether a statistical mixture of the pointer states can be used to build an approximate basis for $\rho_{0}$. 

We plot the lowest nontrivial eigenvalues of $\bar{\mathcal{L}}$ for $n_{at}=\lbrace 1,2,3 \rbrace$ atoms in Fig.~\ref{Fig:liouvillian}. The results are obtained by numerical diagonalization and are sorted in growing order: $\abs{\mathcal{R}e[\lambda_{n}]} < \abs{\mathcal{R}e[\lambda_{n+1}]}$. Panel~\ref{Fig:liouvillian}{(a)} shows the case of a single atom. Here, the smallest eigenvalue $\lambda_{1}$ begins to decrease when the drive is turned on to bring the system out of the ground state. As the drive approaches the collapse point, $\lambda_{1}$ begins to rise until it crosses $\lambda_{2}$. It then takes a sharp dive towards zero that reflects a closing of the gap between $\lambda_{0}$ and $\lambda_{1}$. Afterwards, $\lambda_{1}$ continues to decrease ($\lambda_{1} \to 0$). This trend also appears in the intricate spectra for $n_{at}=2$ and~$3$ atoms, albeit extended to more eigenvalues. The spectra are plotted in panels \figlink{Fig:liouvillian}{(b)} and \figlink{Fig:liouvillian}{(c)}, where it is shown that an initially lowering eigenvalue rises to cross its upper neighbor around the first collapse and then takes a sharp dive to close its gap with $\lambda_{0}$. The behavior repeats around each collapse for the following eigenvalues. 

These two properties of the Liouvillian: level crossing and gap closing, provide a complementing perspective to the behavior found in Secs.~\ref{Sec:cavity_loss} and~\ref{Sec:stochastic}. A level crossing in the Liouvillian spectrum is intimately related to a discontinuity of an eigenoperator~\cite{Kato_2013}. This discontinuity is in agreement with the appearance of multiple distinct peaks in the Wigner distributions around each collapse $\varepsilon_{\rm col}(\abs{m_s})$, as shown in \figref{Fig:multi-peaked}. The closing gaps, in turn, are associated to the slowdown in photon number found in Fig.~\ref{Fig:photon_number}, where an emerging $m_{s}$-ladder is connected by fluctuations to an already existing one. The closing gaps shows how switching between the approximate pointer states of the system written in Eq.~(\ref{eq:pointer-1}) becomes less likely as the drive is increased.

These changes occur gradually and the gap remains open as we are studying systems of finite size. We expect that, as the coupling strength is increased and we move closer to the thermodynamic limit, these changes occur sharply and eventually the gaps fully close.

\subsection{Parity-breaking and Liouvillian spectrum}

The closing gaps support the emergence of multiple eigenoperators $\rho_i$ sharing an eigenvalue zero $\lambda_i=0$ in the thermodynamic limit: $\varepsilon/\kappa \to \infty$, $\Omega/\kappa \to \infty$, and $\varepsilon/\Omega$ constant. This degeneracy is characteristic of a symmetry-breaking phase transition, where each pair $\rho_i$ and $\lambda_i$ belongs to a different symmetry sector~\cite{Minganti_2018, Prosen_2023}. 
The symmetries are obtained from a unitary superoperator $\mathcal{U}$, which commutes with the Liouvillian $\comm{\bar{\mathcal{L}}}{\mathcal{U}}$=0. A well-known symmetry for our system is the $SU(2)$ symmetry associated with the total angular momentum $\hat{S}$. Since the \eqref{eq:liouvillian} does not mix eigenoperators of different total angular momentum, we dropped the $s$ quantum number and reamain within $s=n_{at}/2$. This symmetry is not present when the atoms are separated by individual atomic decays~\cite{Puri_1979,Carmichael_1980,Larson_2018}.

The results shown throughout this work suggest the existence of an additional symmetry that is broken in the thermodynamic limit. This discrete symmetry is related to a sign change in atomic operators $S_{x,y}$ accompanied by one in the field quadratures~$(\hat{a}\pm\hat{a}^{\dagger})$ as found in Ref.~\cite{Curtis_2022}. But, for the few-atom limit explored here, the symmetry is broken in stages. Each collapse~$\varepsilon_{\rm col}(\abs{m_s})$ marks a change in the Liouvillian spectrum where new parity-broken states appear. Before the first collapse, in the empty-cavity phase, we expect a unique steady state that shares the symmetry ($\mathcal{U}\hat{\rho}_{ss} = u_0 \hat{\rho}_{ss}$) of the Liouvillian. After each collapse, in the filled-cavity phase, a mixed steady state composed of the coexisting eigenoperators $\rho_i$ emerges, in accordance with the Wigner distributions plotted above.

The trend that is unveiled in the low-lying Liouvillian spectrum for different atom numbers is indicative of a discrete symmetry similar to the $Z_n$ symmetry-breaking transition predicted for $n$-photon driven nonlinear oscillators in Ref.~\cite{Minganti_2023}. In our case, we find the closing of $2s$ eigenvalues, related to the atomic number as $s=n_{at}/2$. This implies the existence of $2s+1$  different symmetry sectors, which we relate to the number of possible dressed-states ladders in \secref{Sec:cavity_loss}.

\section{Conclusion}\label{Sec:Summary}

Motivated by recent experiments where emitter arrays are strongly coupled to a single mode of a high-Q cavity~\cite{Fink_2009,Liu_2023,Masson_2023}, we have explored cooperative effects in driven dissipative light-matter systems. To do so, we have emphasized the role of the dressed states of the Tavis--Cummings model and shown that they provide a structure to the system that is maintained in the presence of cavity drive and dissipation. 

This underlying structure helps to explaining an organization of the system that is found as the driving field is ramped up. The organization was understood as a dissipative quantum phase transition, as the system displays abrupt changes in macroscopic observables---such as photon number and atomic population---accompanied by structural changes in the Liouvillian spectrum and steady-state probability distributions. It is in this transition that cooperativity becomes apparent. Unlike the transition found for a single-atom~\cite{Carmichael_2015} or mean-field predictions for many emitters~\cite{GJ_2018a}, this few-emitter system displays multiple parity-broken solutions in the ordered phase. Each solution represents an excitation path that is opened along a particular dressed state ladder. Explicit results were presented for small arrays of one to four atoms that, in the ordered phase, were coupled to cavities carrying hundreds of photons to collaborate our findings. 

Separability has been at the center of our research. Here, cavity loss was considered to be the only decay channel. The fact that both coherent and vacuum states are eigenstates of the decay operator $\hat{a}$ was crucial in stabilizing the system above and below the critical drive. Cavity loss was not enough to destroy the correlations that formed between the atomic ensemble and the light field. This allowed us to study the competition between energy-costly fluctuations and  the information exchanged with the environment as the system relaxed towards a steady state. The question of how additional decay channels, such as collective or individual atomic decay, break the correlations and thus help settle this competition remains. 

Additional channels have been studied before in the many-emitter and weak-coupling limits. Collective atomic decay has been shown to lead to collective resonance fluoresence~\cite{Puri_1979,Carmichael_1980,Larson_2018} and individual decay to optical bistability~\cite{Bonifacio_1978, Carmichael_2015}. The question of separability has also risen in the context of superradiance where the coexistence of  individual and collective decay channels requires for atomic correlations to be mantained up to a given order~\cite{Plankensteiner_2022,Oriol_2023}. We hope that exploring these systems in the few-emitter limit will help to gain more theoretical insight towards the role of this competition in the behavior of quantum systems driven far from equilibrium.

\section*{Acknowledgements}

RG-J gratefully acknowledges S.~Masson and A.~Camacho for insightful discussions; and the Applied Quantum Physics Laboratory at Chalmers University of Technology for its hospitality during an extended visit. This project was supported by PAPIIT-UNAM grant IA103024. We acknowledge support from the Knut and Alice Wallenberg Foundation through the Wallenberg Center for Quantum Technology (WACQT). TK and GJ acknowledge support from the Swedish Research Council, VR grant No. 2016-06059 and 2021-04037.


\begin{thebibliography}{1}
\bibitem{us_2023} The dressed state manifolds divide into two kinds depending on the photon number. The first kind occurs for small photon numbers $n < n_{at}/2$ where each manifold contains $n+1$ states. The second kind occurs for all other photon numbers and each manifold has $n_{at}+1$ states.
\end{thebibliography}

\end{document}